\documentclass[12pt]{article}
\usepackage{graphicx}
\usepackage{psfrag}
\usepackage{epsfig} 
\usepackage{psfig} 
\begin{document}
\begin{flushright}
hep-th/0409302\\
\end{flushright}
\vskip 1cm

\begin{center}
{\Large {\bf A note on the  wellposedness of scalar brane world cosmological perturbations}}\\
\vskip 1.5cm
C\'edric Deffayet\footnote{deffayet@iap.fr}\\
{\it GReCO/IAP, FRE 2435-CNRS, \\98bis boulevard Arago, 75014 Paris, France \\
and \\
 F\'ed\'eration de recherche APC, Universit\'e  Paris VII,\\
2 place Jussieu - 75251 Paris Cedex 05, France}
\end{center}
\vskip 1cm
\noindent
\begin{center}
{\bf Abstract}
\end{center}
We discuss scalar brane world cosmological perturbations for a 3-brane world in a maximally symmetric 5D bulk. We show that Mukoyama's master equations leads, for adiabatic perturbations of a perfect fluid on the brane and for scalar field matter on the brane, to a well posed problem despite the {\it non local} aspect of the boundary condition on the brane. We discuss in relation to the wellposedness the way to specify initial data in the bulk.
\vskip .3cm

\pagebreak

\newcommand{\beq}{\begin{eqnarray}}
\newcommand{\eeq}{\end{eqnarray}}
\newcommand{\kc}{\kappa_{(5)}}
\newcommand{\kq}{\kappa_{(4)}}
\newcommand{\kcd}{\kappa^2_{(5)}}
\newcommand{\kcq}{\kappa^4_{(5)}}
\newcommand{\kqd}{\kappa^2_{(4)}}
\newcommand{\Lc}{\Lambda_{(5)}}
\newcommand{\Lq}{\lambda_{(4)}}
\newcommand{\abd}{\dot{a}_{(b)}}
\newcommand{\abp}{a^\prime_{(b)}}
\newcommand{\ab}{a_{(b)}}
\newcommand{\nbd}{\dot{n}_{(b)}}
\newcommand{\nbp}{n^\prime_{(b)}}
\newcommand{\nb}{n_{(b)}}
\newcommand{\bb}{{(b)}}
\newcommand{\MM}{{(M)}}
\newcommand{\EE}{{({\cal E})}}
\newcommand{\hd}{\dot{H}}
\newcommand{\hdd}{\ddot{H}}
\newcommand{\hddd} {H^{\cdot \cdot \cdot}}
\newcommand{\da}{\dot{a}}
\newcommand{\db}{\dot{b}}
\newcommand{\dn}{\dot{n}}
\newcommand{\dda}{\ddot{a}}
\newcommand{\ddb}{\ddot{b}}
\newcommand{\ddn}{\ddot{n}}
\newcommand{\paBDL}{a^{\prime}}
\newcommand{\pb}{b^{\prime}}
\newcommand{\pn}{n^{\prime}}
\newcommand{\ppa}{a^{\prime \prime}}
\newcommand{\ppb}{b^{\prime \prime}}
\newcommand{\ppn}{n^{\prime \prime}}
\newcommand{\fda}{\frac{\da}{a}}
\newcommand{\fdb}{\frac{\db}{b}}
\newcommand{\fdn}{\frac{\dn}{n}}
\newcommand{\fdda}{\frac{\dda}{a}}
\newcommand{\fddb}{\frac{\ddb}{b}}
\newcommand{\fddn}{\frac{\ddn}{n}}
\newcommand{\fpa}{\frac{\paBDL}{a}}
\newcommand{\fpb}{\frac{\pb}{b}}
\newcommand{\fpn}{\frac{\pn}{n}}
\newcommand{\fppa}{\frac{\ppa}{a}}
\newcommand{\fppb}{\frac{\ppb}{b}}
\newcommand{\fppn}{\frac{\ppn}{n}}
\newcommand{\UU}{\Upsilon}
\newcommand{\fU}{f_\UU}
\newcommand{\CC}{{\cal C}}

\section{Introduction}
Brane world models \cite{Arkani-Hamed:1998rs,Randall:1999vf,DGP} lead generically to modifications of the gravitational interaction which get reflected into cosmology (see e.g. \cite{BDL,Deffayet:2001uy}). Aside from the fact this can potentially lead to cosmological signatures of extra-dimensions, this has also been exploited both for phenomenological constructions and to study more abstract questions.  On this ``abstract'' side, one can mention works done in the framework of the 
 {\it brane induced gravity} model of Dvali Gabadadze and Porrati (DGP model in the following) \cite{DGP}. This model can serve as a toy  (see e.g. \cite{TOYDGP})  for exploring some long standing problems related to massive gravity, such as the van Dam-Veltman-Zakharov (vDVZ) discontinuity \cite{Veltman}. It  has also the virtue of leading to interesting cosmological consequences for the recent universe \cite{Deffayet:2001uy,acc}, although the phenomenological viability of such a scenario 
is currently subject to a debate \cite{STRONG}. As far as observational consequences of brane world cosmology  are concerned, there is currently  no firm predictions of the possible signatures  of these scenario  in the CMB or Large Scale Structures.  This is due in part to the fact that cosmological perturbations 
on brane worlds are very poorly understood despite the large number of works dealing with them and some recent progresses \cite{KOYA}.
Brane world cosmological perturbations are also interesting to study in relation with the vDVZ discontinuity \cite{Tokyo}. 
It is some aspects of those  perturbations we would like to address in this
work.

 We will concentrate on the case of a flat 3-brane of codimension one, and  scalar perturbations (i.e., scalar from a brane observer point of view).
One can first use a point of view relying on a 4D projection of 5D Einstein's equations  on the brane \cite{Shiromizu:2000wj}. 
In this approach the influence of the bulk is encoded into additional degrees of freedom with respect to the usual 4D ones, the so-called Weyl fluid \cite{Langlois,LMSW,Bridgman:2001mc,Roy1,Langlois:2001ph}. 
Knowing the time evolution of those Weyl degrees of freedom is however required to know the one of standard cosmological perturbations, but not all 
 the Weyl d.o.f. have a local evolution equation on the brane. This means that one has in general to solve the full 5D problem in the bulk,  and provide some initial conditions there to know the evolution of the brane perturbations. In the simplest case of a maximally symmetric bulk, one can show that the scalar perturbed Einstein equations  are solved by a single master variable verifying a second order hyperbolic partial differential equation (PDE), the master equation \cite{Mukohyama:2000ui}. The matter equation of state, in the simplest case of adiabatic perturbations
of a perfect fluid, translates into a boundary  condition on the brane for the master variable \cite{Kodama:2000fa,Deffayet:2002fn}. This boundary condition is ``non local'' in the sense that it contains derivatives with respect to the brane cosmic time of the master variable and of its normal derivative with respect to the brane\footnote{We do not find very well suited the terminology ``non-local'' to describe the boundary condition of interest here, since it only contains a finite number of derivatives, and, as shown in this note, the problem can be recast into a standard hyperbolic problem. The ``non locality'' of the problem (as seen by a brane observer) only comes from the fact that the brane is embedded into a bulk from where is can receive informations, and not from the form of the boundary condition. The latter is indeed ``local'' when one considers vectors or tensors perturbations \cite{Kodama:2000fa}. We will however stick with this terminology to agree with the literature.}. It is thus not of the usual linear combination of Neumann and Dirichlet form.
We will show in this work that the same holds true for the case where the only matter on the brane consists in  a scalar field.
The main purpose of this note is to show that the differential problem, associated with the master equation formalism, is  well posed in general,
despite the non local aspect of the boundary condition on the brane.
We will also discuss correlatively the initial conditions to be provided. 
The proof is quite straightforward, and is based on a rewriting of the master equation as a PDE of order higher than two in such way the boundary condition on the brane takes a standard form. We feel however that 
a clear presentation of such a proof can help a numerical implementation of this problem, but also provides a way, different from the usual one, to look at the scalar brane world cosmological perturbations as deriving from a standard (``local'', that is to say with no derivatives along the brane) well-posed hyperbolic problem. This matters in particular when addressing issues such as stability of brane worlds \cite{Ringeval:2003na}. Last, the wellposedness does also play a r\^ole when using cosmological perturbations of DGP model to study the vDVZ discontinuity on 
 cosmological backgrounds \cite{Tokyo}. 

The paper is organized as follows. In the next section, we introduce the differential problem associated with brane world scalar cosmological perturbations in the Master equation formalism. We then show how this problem can be recast in a well posed form.

\section{The differential problem associated with brane world scalar cosmological perturbations}
\subsection{Background cosmological solutions} \label{BACK}
We consider models where our usual 4D space-time is the worldvolume of a 3-brane in a 5D bulk space-time. The latter, when not perturbed, 
will be taken to be a slice of a maximally symmetric space-time. In  addition we will not consider any stabilization mechanism, so that the bulk will be considered empty, and moreover we will assume a $Z_2$ symmetric brane. 
This setup is well suited  for e.g. the Randall-Sundrum II model 
\cite{Randall:1999vf} ({\it RS model} in the following) where the (unperturbed) bulk is a slice of $AdS_5$ or for the Dvali-Gabadadze-Porrati model \cite{DGP}, where the bulk is a slice of 5D Minkowski space-time. It is also the simplest setting that can be considered in order to deal in an exact (and controllable)  way with homogeneous and inhomogeneous cosmology in brane worlds. 
The line element of the homogeneous cosmological solution reads, in a so-called 
Gaussian Normal Coordinate system (GN in the following),
\beq \label{backmet}
ds^2_{(5)} = -n^2(t,y) dt^2 + a^2(t,y) \delta_{ij} dx^i dx^j + dy^2, 
\eeq
where 
the exact form of the functions $a$ and $n$, which will not be used explicitly here, is model dependent, and has been worked out in the literature \cite{Kaloper,BDL}. In this coordinate system\footnote{Note that this coordinate system
 is usually breaking down somewhere in the bulk, where the functions $a$ or $n$ are vanishing. However the bulk space-time of interest here being non singular, it is always possible to choose a better coordinate system for the bulk. Aspects of this issue are discussed at the end of the next subsection.}, the brane is the $y=0$ hypersurface,
and the induced metric on the brane is FLRW, so that $x^i$ can be considered as  comoving coordinates of standard cosmological observers on the brane. One can in addition choose the time parametrization so that $n=1$ on the brane, and such that $t$ is the standard cosmological time.  The function  $a(t,0) \equiv a_\bb(t)$ is then the scale factor  of the FLRW 
metric on the brane.
 
As far as homogeneous cosmology is concerned, the whole space-time can be 
thought of as a slice of a 5D maximally symmetric space-time glued to a copy of itself along the brane, the latter being a 
 FLRW space-time. This generalizes (see e.g. \cite{Deruelle:2000ge,Deffayet:2001uy})  the well known  global picture \cite{Gibbons:1993in} of a domain wall space-time  where the metric on the wall (i.e., the brane) is de Sitter \cite{domainwall}. 
We now turn to discuss  cosmological perturbations.

\subsection{Scalar cosmological perturbations and the master equation}
To deal with cosmological perturbations in the models of interest, one can use \cite{Kodama:2000fa,Mukohyama:2000ui,Langlois} the standard {\it scalar, vector, tensor} decomposition (see e.g. \cite{Mukhanov:1992me}), depending on the way the respective perturbations transform under the isometries of the 3D maximally symmetric slices of  metric (\ref{backmet}). We will only consider in this article the case of scalar cosmological perturbations, which is the most difficult one, but also the most interesting as far as phenomenology is concerned.
 We summarize, in this subsection,  some results obtained elsewhere on scalar brane world cosmological perturbations. 

It has been shown by Mukohyama \cite{Mukohyama:2000ui} (see also \cite{Kodama:2000fa}), that  the linearized scalar  Einstein's equations over a
  maximally symmetric bulk can be conveniently
solved introducing a master variable $\Omega$
 which obeys a PDE in the bulk, the master equation.
The latter, when $\Omega$ has a non-trivial  dependence  in the comoving coordinates $x^i$ and when the brane has flat spatial sections, reads in a  GN coordinate system (\ref{backmet})

\beq 
\label{FQ1}
\left(\frac{{\Omega^\cdot}}{na^3}\right)^\cdot + \left(\frac{\Lc}{6}-
\frac{\Delta}{a^2}\right) \frac{n \Omega}{a^3} - \left(\frac{n
\Omega^\prime}{a^3} \right)^\prime=0,
\eeq
where $\Delta$ is defined by $\Delta = \delta^{ij} \partial_i \partial_j$, a dot denotes a derivative with respect to $t$, a prime a derivative with respect to $y$, and $\Lambda_{(5)}$ is the bulk cosmological constant. 
In the rest of this article, we will implicitly consider all the perturbations as Fourier transformed  
with respect to the $x^i$'s, so that the perturbations will all be functions of two coordinates ($y$ and $t$ or the characteristic coordinates introduced below, X and Y) and a comoving momentum $\vec{k}$. At the linear level of this work, the different modes do not mix, and one can do a mode by mode analysis.
In particular, replacing $\Delta$  by  $-\vec{k}^2$ in equation (\ref{FQ1}), 
 the master equation takes the form of a second order ($-\vec{k}^2$ 
 dependent) hyperbolic differential operator acting on $y$ and $t$ dependent functions. In the following, we will also use characteristic coordinates $X$ and $Y$ for this operator, such that the master equation (\ref{FQ1}) reads  
\beq \label{MASTERXY}
\partial_{X}\partial_{Y} \Omega = U_\Delta(X,Y) \partial_X \Omega + V_\Delta(X,Y) \partial_Y \Omega + W_\Delta(X,Y) \Omega,
\eeq where the coefficients $U_\Delta,V_\Delta,W_\Delta$ and the unknown function $\Omega$ are functions of $X$ and $Y$ (and also of the wave number $\vec{k}$).
For example, when the bulk is a slice of 5D Minkowski, the master equation reads in 
characteristic coordinates $(X,Y)$ \cite{Deffayet:2002fn}
\beq
 \partial_{X} \partial_Y  \Omega = \frac{3 \partial_Y \Omega}{2 X} + \frac{\Delta \Omega}{4 X^2}. \label{nullDGP}
\eeq
In those coordinates,  the bulk metric reads 
\beq \label{MINKOCOS}
ds^2 = -dX dY + X^2 dx^i dx^j \delta_{ij},
\eeq
\label{PCGI22p18}where $X$ is given by $a(t,y)$ and $Y$ is a function of the Gaussian normal coordinates 
$y$ and $t$ which can be easily obtained e.g. by the coordinate change given in \cite{Deruelle:2000ge}. 

\begin{figure}
\centering
\psfrag{C}{\large brane}
\psfrag{f}{\large cosmic time}
\psfrag{c}{\large initial curve ${\cal C}_{(I)}$}
\psfrag{T}{\large Y}
\psfrag{d}{\large domain ${\cal D}$}
\psfrag{A}{\large O}
\psfrag{H}{\large E}
\psfrag{t}{\large X}
\psfrag{e}{\large C}
\psfrag{x}{\large X}
\psfrag{i}{\large initial time}
\psfrag{b}{\large brane ${\cal C}_\bb$}
\resizebox{14cm}{7cm}{\includegraphics{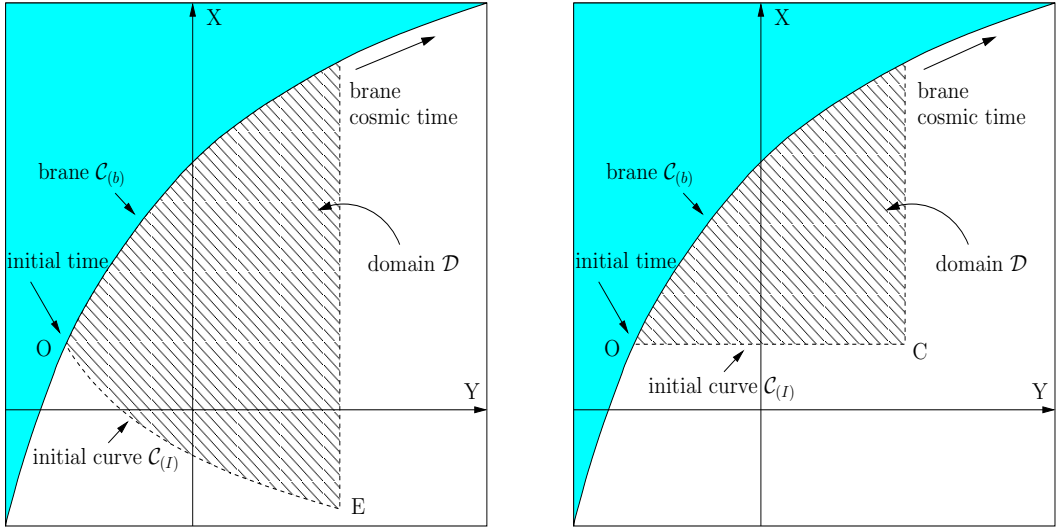}}
\caption{Schematic representation of the bulk space-time with the
  brane trajectory, ${\cal C}_\bb$, in the characteristic $(X,Y)$ coordinates. 
The gray (cyan) part is cutoff and the complete bulk space-time is made out of the remaining part, glued to a copy of itself along the brane.  If a given initial event, O,  is chosen along the brane from which one wishes to evolve cosmological perturbations, 
one needs to specify a boundary condition along the brane, and
initial data in the bulk along an initial curve ${\cal C}_{(I)}$. We will 
discuss  two different cases, first the case where the initial curve is non characteristic (left figure), second the case where the initial curve is characteristic (right figure). When initial data are specified along [OE[ (resp. [OC[), and a boundary condition is supplied along the brane, the solution for the master equation can be determined in the (hatched) {\it domain of determinacy} ${\cal D}$.} 
\label{fig2}
\end{figure}

The (background) brane trajectory gives a boundary,  ${\cal C}_\bb$, 
in the characteristic coordinates $(X,Y)$ plane of the domain, ${\cal D}$,  where we want to solve the master equation.
Indeed, as recalled in the following section, the master variable obeys a boundary condition on ${\cal C}_\bb$, obtained from the brane matter equation of state. The boundary of  ${\cal D}$ is thus defined by ${\cal C}_\bb$ but also by some curve, denoted ${\cal C}_{(I)}$, where we need to specify initial data. This curve can either be nowhere characteristic, but also have some characteristic part. We call $O$ the intersection point of    ${\cal C}_\bb$ and  ${\cal C}_{(I)}$ in the $(X,Y)$ plane (see figure \ref{fig2}).

The curve ${\cal C}_\bb$ is  
 defined by two functions $X_{(b)}(t)$ and $Y_{(b)}(t)$ ($t$ being the cosmological time of a standard cosmological observer on the brane) that can be obtained e.g. from the relation between characteristic coordinates and Gaussian Normal coordinate as well as from  the cosmological solutions referred to in subsection \ref{BACK}.
 The exact form of those functions depends on the model considered and in particular on the matter content of the brane through the brane Friedmann's equations.  However, as far as the conclusions of this paper are concerned, the only important point is  that those functions can be both taken as strictly increasing functions of the cosmic time, and in addition, are such that the brane trajectory is nowhere (except possibly at the Big Bang, see below)  characteristic (meaning here that the brane trajectory is nowhere tangent to one of characteristic direction). This holds true for  trajectories of phenomenological interest. For example, for a brane in Minkowski bulk, as 
is the case for the Dvali-Gabadadze-Porrati model,  the brane trajectory is given by 
\beq
X_\bb &=& a_\bb,  \nonumber\\
\dot{Y}_\bb &=& {\dot{a}^{-1}_\bb} \nonumber.
\eeq 
It is then obvious to verify that this trajectory obeys the conditions enumerated above as long as one has $\dot{a}_\bb >0$. 
Moreover, if one assumes that the scale factor on the brane, $a_\bb$, behaves as $t^\alpha$ close to $t=0$,  one finds that (for $0<\alpha<2$, which is the case for e.g. radiation or matter dominated standard or brane cosmology) $Y$ converges as $t$ goes to zero, so that the Big-Bang is indeed in this case a point in the two dimensional $(X,Y)$ plane.  
The coordinate system (\ref{MINKOCOS}) however breaks down 
in $X=0$. A closer look at the way the brane is embedded in the bulk, indeed shows \cite{Ishihara:2001qe,Deffayet:2002fn,Lue:2002fe}  that the brane is a conoidal 4D surface with a summit, an event in 5D Minkowski space-time,
 which is the Big Bang. The brane is also tangent to the light cone of the Big Bang along a line which corresponds to the $X=0$ coordinate singularity. However, one can also go to any point along this line, staying on the brane, when $t$ goes to zero, so that the Big Bang also extends along this line. 
A similar picture holds true for the case of an $AdS_5$ bulk (see \cite{Deffayet:2002fn} for a more extended discussion of this and of the case where the brane has no Big Bang). 
\label{WELLPOSED 12-2}

\subsection{Boundary condition on the brane from matter equation of state}
So far we have only discussed the bulk perturbed Einstein's equations, as well as the background (unperturbed) brane trajectory. We now turn to perturbations of the brane induced metric, as well as those of brane localized matter. We discuss  how to obtain a boundary condition on the brane, from the matter equation of state in simple cases of interest.
After an appropriate gauge choice, the linearized induced metric on the brane can be put in the 4D longitudinal form, which reads (keeping only the scalar part)
\beq
ds^2 = -(1+ 2 \Phi) dt^2 + a^2_\bb(t)(1-2 \Psi) \delta_{ij} dx^i dx^j,
\eeq
where $a_\bb$ is the background scale factor on the brane,  $\Phi$ and $\Psi$ are metric perturbations. 
The scalar perturbations, $\delta T^\mu_\nu$ of the matter energy-momentum tensor $T^\mu_\nu$,  are decomposed as 
\beq
\delta T^0_0 &=& -\delta \rho, \label {dT00}\\ 
\delta T^0_i &=& \partial_i \delta q,  \label{dT0i} \\
\delta T^i_j &=& \delta P \delta^i_j +  \left(\delta^{ik} \partial_k \partial_j-  \frac{1}{3}\delta^i_j \Delta  \right) \delta \pi. \label{dTij}  
\eeq
From the point of view of an observer located on the brane, the functions 
$\Phi$, $\Psi$, $\delta \rho$, $\delta q$, $\delta P$, $\delta \pi$ are those 
 which time evolution she would like to know to compare with standard 4D results. These functions are all expressible in terms of the master variable $\Omega$ and the anisotropic stress $\delta \pi$.  If one further assumes that  the anisotropic stress of matter vanishes (this would hold true for matter on the brane consisting only in a perfect fluid or a scalar field, those two cases will be discussed in the following), one gets expressions for all the induced metric and matter perturbations on the brane in terms of the master variable $\Omega$
in the form \cite{Deffayet:2002fn}
\beq
\Phi = \sum_{s=0}^{s=s_{(0)}^\Phi} \aleph^\Phi_{(0,s)} \partial^s_t \Omega 
+ \sum_{s=0}^{s=s_{(1)}^\Phi} \aleph^\Phi_{(1,s)} \partial^s_t \Omega^\prime, \label{EXPPERT}
\eeq
with similar expressions for $\Psi, \delta \rho, \delta q, \delta P$. The integers $s_{(1)}$ and $s_{(0)}$, as well as 
the coefficients $\aleph$ are model dependent. 
The latter coefficients are functions of the brane background induced metric (thus of the brane cosmic time) and the 3-momentum square $\vec{k}^2$. They have been given in \cite{Deffayet:2002fn} for RS and DGP models.
Expressions (\ref{EXPPERT}) are not {\it per se} giving a boundary condition on the brane for $\Omega$, since their left hand side is not known as a function of time. However the sought for boundary condition can be obtained from  the matter ``equation of state'', as we now explain.

The simplest case one can imagine is the case of an adiabatic equation of state, for which one has 
\beq \label{stateADIAB}
\delta P = c_{S}^2 \delta \rho,
\eeq
where $c_{S}$ is the sound speed. This relation translates into a boundary condition on the brane for the 
master variable $\Omega$ once one replaces $\delta P$ and $\delta \rho$ in the above equation by their expressions as a function of $\Omega$ and $\Omega^\prime$ of the form (\ref{EXPPERT}). This boundary condition can be written as 
\beq \label{POLYBOUND} 
\Pi_0 (\Omega) + \Pi_1(\Omega^\prime) = 0, 
\eeq
where  $\Pi_0$ and $\Pi_1$ are polynomials in cosmic time derivative $\partial_t$, with cosmic time (and $\vec{k}^2$) dependent coefficients. Their expression can be deduced from the expressions of 
$\aleph^P$ and $\aleph^\rho$ given in \cite{Deffayet:2002fn}. For the RS and DGP models, the degrees of $\Pi_0$ and $\Pi_1$ are strictly larger than 1, so that the boundary condition (\ref{POLYBOUND}) is not of the usual linear combination of Neumann and Dirichlet form.

Another example leading to a similar boundary condition is 
the case where the only matter present on the brane is a  scalar field. In this case the anisotropic stress $\delta \pi$ also vanishes, but equation 
(\ref{stateADIAB}) no longer holds true. Instead one now has 
\beq \label{statescalar}
\delta \rho -\delta P  - \delta q\left( 2 \frac{\ddot{\phi^B}}{\dot{\phi^B}} + 6 H \right) =0,
\eeq
where $\delta \rho$, $\delta P$ and $\delta q$ are defined as in equations 
(\ref{dT00}-\ref{dTij}), with $T^\mu_\nu$ being the energy momentum tensor of the (canonically normalized) scalar field (with an arbitrary potential), and $\phi^B$ the background value of the scalar field. This equation can easily
 be obtained from the form of the energy momentum tensor of the scalar field, as well as from the conservation equation of its energy momentum tensor. It does not require knowledge of the gravitational dynamics (i.e., 4D Einstein's equation need not be valid) and can be considered as an equation of state for the scalar field.  Following the same path as above and replacing $\delta \rho$, $\delta P$ and $\delta q$ by their expressions as functions of the master variable $\Omega$, one gets a boundary condition on the brane of a similar form as equation (\ref{POLYBOUND}). 

Note that it is only in very special cases, such as the ones studied here, that a boundary condition of the form (\ref{POLYBOUND}) can be found from the brane matter equations of motion (or equation of state). In the most general cases, one do not expect to be able to extract from those equations a simple boundary condition for the master variable. One should instead solve simultaneously (i.e. numerically) those equations and the hyperbolic bulk problem. 

In the next section, we show that despite the {\it  non local} form of the boundary condition (\ref{POLYBOUND}), the differential problem associated with 
brane world scalar cosmological perturbations can be recast into a standard hyperbolic form (including the boundary condition), and is well posed once one provides suitable initial data in the bulk.

\section{Wellposedness}
\subsection{Bulk equation and brane boundary condition}
The peculiar features of the boundary condition (\ref{POLYBOUND}) are that it involves derivatives of the master variable of order strictly larger than one
as well as derivatives of it along the boundary curve, while the PDE in the bulk is of order two. This holds for 
both cases mentioned above, namely boundary conditions derived from equations of state (\ref{stateADIAB}) and (\ref{statescalar}). In the RS model, indeed the highest derivative of $\Omega$ appearing in the expression of 
$\delta \rho$, $\delta q$ and $\delta P$ is $\Omega^{\cdot \cdot \prime}$ \cite{Deffayet:2002fn}, which appears in $\delta P$. 
This terms appears with a coefficient $\aleph^{\delta P}_{(1,2)} = (\kappa_{(5)}^2 a_{(b)})^{-1}$ (where $\kappa_{(5)}^2$ is the inverse cube of the 5D reduced Planck mass). This coefficient never vanishes, while  all other coefficients $\aleph_{(r,s)}^{\delta P}$, $\aleph_{(r,s)}^{\delta \rho}$ and $\aleph_{(r,s)}^{\delta q}$ vanish for $r+s \geq 3$ (as can be seen from expressions given in \cite{Deffayet:2002fn}). Furthermore, one can decompose  $\Omega^{\cdot \cdot \prime}$ as 
\beq
\Omega^{\cdot \cdot \prime} = \left(\partial_tX\right)^2\left(\partial_yX\right)\partial_X^3 \Omega +  \left(\partial_tY\right)^2\left(\partial_yY\right)\partial_Y^3 \Omega
+ ...\eeq
where the dots mean terms proportional to  partial derivatives $\partial^r_X\partial^s_Y\Omega$, with $r\leq 2$, $s\leq 2$, and $r+s \leq 3$. Because the brane trajectory is nowhere characteristic, $\partial_tX$ and $\partial_yX$ do not vanish on the brane (and similarly $\partial_tY$ and $\partial_yY$). 
One can deduce from the above discussion that one can 
 rewrite (\ref{POLYBOUND}), for the RS model, as  
\beq \label{BOUNDARY}
\partial^{r_{\rm max}}_X \Omega = \sum_{ r+s\leq r_{\rm max}} \alpha_{(r,s)} \partial^r_X \partial^s_Y \Omega,
\eeq
 where $r_{\rm max}=3$, 
$\alpha_{(r_{\rm max},0)}$ vanishes by definition and the others $\alpha_{(r,s)}$ are known functions of the parameter $t$ along the brane (the cosmic time). 
 A similar rewriting (with $r_{\rm max}=4$) is possible in the case of the DGP model. 
In this case, the derivative of $\Omega$ of highest order appearing in the expression of $\delta P$ is $\Omega^{\cdot \cdot \cdot \cdot}$, and there is no other non vanishing fourth order derivatives in $\delta \rho$,  $\delta q$ and $\delta P$. The coefficient $\aleph_{(0,4)}^{\delta P}$ is given by 
\cite{Deffayet:2002fn}
\beq
\aleph_{(0,4)}^{\delta P} = -\frac{2 + \Upsilon}{2 a_\bb \kappa_{(4)}^2 \left(\Upsilon + \frac{2 H^2 +\dot{H}}{H^2}\right)}.
\eeq
In the above expression $\kappa_{(4)}^2$, is the inverse square of the 4D reduced Planck mass, $\Upsilon$ is given by $\Upsilon = \epsilon / H r_c$, where $\epsilon = \pm 1$ depending on the branch of solutions considered for the background, and $r_c$ is a distance scale parameter characterizing the transition from 4D to 5D gravity in the DGP model
 (see \cite{Deffayet:2002fn} for more details). 
The coefficient $\aleph_{(0,4)}^{\delta P}$ does not vanish for physically interesting cases, so that (\ref{POLYBOUND}) can indeed be recast into the form 
(\ref{BOUNDARY}) (with $r_{\rm max}=4$). The form of the boundary condition (\ref{BOUNDARY}) leads us to define the  
$(r_{\rm max}+1)(r_{\rm max}+2)/2$ unknowns $u_{(r,s)}$, 
with $0 \leq r \leq r_{\rm max} $, $0 \leq  s \leq r_{\rm max}$ and $r+s \leq r_{\rm max}$ by 
\beq
u_{(r,s)}= \partial^r_X \partial^s_Y \Omega, 
\eeq
and to recast the differential equation (\ref{MASTERXY}) in the following linear system (for $r_{\rm max} \geq 1$) 
\beq \label{equa1}
\partial_X u_{(r,s)} &=& u_{(r+1,s)} ,\\&& 
 \quad {\rm for} \quad r+s < r_{max}, \nonumber \\ \label{equa2}
\partial_X u_{(r,s)} &=& \partial^{r}_X \partial^{s-1}_Y \left( U_\Delta u_{(1,0)} + V_\Delta u_{(0,1)} + W_\Delta  u_{(0,0)} \right) ,\label{deux}  \\
&&\quad {\rm for} \quad r+s = r_{max} \quad {\rm and} \quad s \geq 1, \quad {\rm and} \nonumber\\ \label{equa3}
 \partial_Y u_{(r_{\rm max},0)} &=& \partial^{r_{\rm max}-1}_X  \left( U_\Delta  u_{(1,0)} + V_\Delta u_{(0,1)} + W_\Delta  u_{(0,0)} \right). \label{trois}
\eeq
Note that to obtain equations (\ref{deux}) and (\ref{trois}), we have derived the master equation (\ref{MASTERXY}) with respect to the characteristic coordinates $X$ and $Y$, so that, out of a single second order PDE, we are  making 
up a system of PDEs of order higher than two (here written in equations (\ref{equa1}-\ref{equa3}) as a linear first order system, in its {\it characteristic normal form}, see e.g. \cite{courant}), for which the boundary condition (\ref{BOUNDARY}) takes a standard form. The latter reads indeed
\beq \label {BOUNTY}
u_{(r_{\rm max},0)} = \sum_{r+s\leq r_{\rm max}} \alpha_{(r,s)} u_{(r,s)},
\eeq
with $\alpha_{(r_{\rm max},0)} =0$.
The characteristics of the first order linear system (\ref{equa1}--\ref{equa3})
are one family of $X =\; constant$ lines, and $(r_{\rm max}+1)(r_{\rm max}+2)/2-1$ identical families of $Y=\; constant$ lines. Looking at the domain of interest (see figure \ref{fig2}), one sees that, in the vicinity of O, there is only one ingoing characteristic in the domain ${\cal D}$, the $X= $ constant line, for one boundary condition (\ref{BOUNTY}). Standard theorems on linear differential systems then insure that the differential problem associated with the system (\ref{equa1}--\ref{equa3}) and the boundary condition (\ref{BOUNTY}), is well posed 
provided one gives initial data of Cauchy type on the initial curve ${\cal C}_{(I)}$, that is to say the values there of all the variables $u_{(r,s)}$. 
Under this condition, the problem has a unique solution in the domain ${\cal D}$ (limited by the brane and the initial curve), and this solution depends continuously on any parameter  which initial and boundary data may depend on\footnote{For this statement to hold, one has to assume some regularity properties on the coefficients of the differential equation ${\cal L}$, namely Lipschitz continuity \cite{courant}. This would however naturally hold true as long as the coefficients are continuously differentiable, and that one restrict the domain ${\cal D}$ to some compact subdomain. Note further that the wellposedness does not imply the absence of singularities in the solution}. The specification of the initial values for 
 $u_{(r,s)}$ requires however some care, as we now discuss.

\subsection{Bulk initial data}
 Given the fact that the master equation in the bulk is of order two, it is indeed not possible to choose freely on ${\cal C}_{(I)}$ all the values of  
the $u_{(r,s)}$ without encountering inconsistencies. We will consider separately two different cases, first the case where the initial curve ${\cal C}_{(I)}$ is 
not characteristic in O (that is to say that its tangent in $O$ is not parallel to the Y axis), and secondly the case where the initial curve is a characteristic line ($X = \; constant$). We will ask in both cases what freedom we have upon the specification of the initial values of the 
$u_{(r,s)}$ on ${\cal C}_{(I)}$.

If the initial curve ${\cal C}_{(I)}$ is non characteristic in O, it will remain so by continuity in some open neighborhood of O, say [OE[ (see figure \ref{fig2}). In this case, one can specify freely on [OE[ the value of $\Omega$ and of its derivative normal to ${\cal C}_{(I)}$ (those are Cauchy data on the Cauchy curve ${\cal C}_{(I)}$ for the second order differential operator defined by the master equation (\ref{MASTERXY})). From this specification, the initial values of $u_{(0,0)}$, $u_{(1,0)}$ and $u_{(0,1)}$ are known on [OE[. Because $\Omega$ obeys a second order hyperbolic PDE, the  knowledge of $\Omega$ and the normal derivative of $\Omega$ along [OE[ is then enough to compute the values of all the higher order derivatives, $u_{(r,s)}$ with $r+s \geq 2$,  along $[OE[$. This follows immediately from the fact that  [OE[ is not characteristic (and is indeed the definition of a non characteristic line). 

We then turn to the case where the initial curve 
is parallel to the Y axis in some open neighborhood 
[OC[ of O (see figure \ref{fig2}).  In this case, as is well known, one cannot freely specify both $\Omega$ and its normal derivative 
on [OC[. Indeed,  the PDE (\ref{MASTERXY})
can be integrated along a $X = \;constant$ segment on which $\Omega$ is known, to give the normal derivative $\partial_X \Omega$, on this segment, provided 
the value of $\partial_X \Omega$ is known at one point along 
[OC[, e.g. in O.  This can be generalized by recursion to higher order derivatives. 
 Namely if one assumes that one knows along  [OC[ the values of all the derivatives of order lower or equal to a given $m$, with $m\geq 1$, one first gets from those values, and the differentiation of the master equation  (\ref{MASTERXY}), the values of all derivatives of the form $\partial^r_X \partial^s_Y \Omega$ with $s \geq 1$ and $r+s=m+1$. One can however  freely specify the value at O of the derivative $\partial^{m+1}_X \Omega$. This provide an initial value for an ordinary differential equation along [OC[ for the unknown function $\partial^{m+1}_X \Omega$ that one can get by differentiating 
(\ref{MASTERXY}) $m$ times with respect to $X$. The latter can then be integrated to yield $\partial^{m+1}_X \Omega$ all along [OC[. This freedom to specify the derivatives 
$\partial^r_X \Omega$ in O is reminiscent of the wellposedness of the  characteristic initial problem, for a second order PDE such as (\ref{MASTERXY}), where one specifies values of $\Omega$ along two intersecting characteristics, $X=\; constant$ and $Y=\;constant$. To summarize, in this case, one can specify freely the value of $\Omega$ all along the initial segment 
[OC[, as well as the values of the derivatives $\partial^{r}_X \Omega$, for $r \geq 1$ at O. If one does so the values of the $u_{(r,s)}$ will be known  along the initial curve [OC[.  

Note that in both cases described above, one should make sure that the chosen initial data are consistent with the boundary condition (\ref{BOUNTY}).
Otherwise the solution for $\Omega$ will have discontinuities in derivatives of order $r_{\rm max}$ (see e.g. \cite{courant}). Note also that once the  initial values of $u_{(r,s)}$ are chosen on ${\cal C}_{(I)}$, one can compute from the expressions (\ref{EXPPERT}) the initial values on the brane (that is to say in O) of the quantities $\Phi$, $\Psi$, $\delta \rho$, $\delta q,$ and $\delta P$ of interest for an observer on the brane. 

\vspace{1cm}
To summarize and conclude, we have shown here that despite the non local aspect of the boundary condition on the brane for scalar cosmological perturbations of a perfect fluid with adiabatic perturbations or of a scalar field, the differential problem associated with Mukohyama's master equation is well posed and can be recast in a standard hyperbolic problem, as far as the boundary condition and PDE is concerned. Care has however to be taken on the specification of the initial data in the bulk, in order to maintain consistency, as explained hereabove.

\section*{Acknowledgments}
We thank G.~Dvali, K.~Koyama, D.~Langlois, A.~Lue, K.~Malik, M.~Porrati, R.~Scoccimarro, J.~Shata, T.~Tanaka and M.~Zaldarriaga for useful discussions, as well as G.~Esposito-Farese for his precious help to improve the form of this article.

\end{document}